# Monotonic and Non-Monotonic Solution Concepts for Generalized Circuits

Steffen Schuldenzucker
University of Zurich
schuldenzucker@ifi.uzh.ch

Sven Seuken
University of Zurich
seuken@ifi.uzh.ch

July 30, 2019

**Abstract**

Generalized circuits are an important tool in the study of the computational complexity of equilibrium approximation problems. However, in this paper, we reveal that they have a conceptual flaw, namely that the solution concept is not *monotonic*. By this we mean that if $\varepsilon < \varepsilon'$, then an $\varepsilon$-approximate solution for a certain generalized circuit is not necessarily also an $\varepsilon'$-approximate solution. The reason for this non-monotonicity is the way Boolean operations are modeled. We illustrate that non-monotonicity creates subtle technical issues in prior work that require intricate additional arguments to circumvent. To eliminate this problem, we show that the Boolean gates are a redundant feature: one can simulate stronger, monotonic versions of the Boolean gates using the other gate types. Arguing at the level of these stronger Boolean gates eliminates all of the aforementioned issues in a natural way. We hope that our results will enable new studies of sub-classes of generalized circuits and enabler simpler and more natural reductions from generalized circuits to other equilibrium search problems.

## 1 Introduction

Generalized circuits (Chen, Deng and Teng, 2009) have become a vital tool in the study of the computational complexity of equilibrium approximation problems. Reductions from generalized circuits have been used to show PPAD-completeness of a wide range of such problems, including the approximate search problems for: Nash equilibrium of a normal-form game (Daskalakis, Goldberg and Papadimitriou, 2009; Chen, Deng and Teng, 2009; Daskalakis, 2013; Babichenko, Papadimitriou and Rubinstein, 2016; Rubinstein, 2018), Arrow-Debreu market equilibrium (Chen, Paparas and Yannakakis, 2017), competitive equilibrium from equal incomes (Othman, Papadimitriou and Rubinstein, 2016), and payment equilibrium in a financial network (Schuldenzucker, Seuken and Battiston, 2017, 2019).

A generalized circuit consists of *nodes* that are connected by *gates*. Nodes take values between 0 and 1 and each gate defines a constraint on the values of the nodes connected to it. Generalized circuits differ from regular algebraic circuits in three important aspects. First, in addition to arithmetic gates (constants, addition, subtraction, and scaling by a constant), there are also a *comparison gate* and *Boolean gates* that implement the standard Boolean operations (AND, OR, NOT). Second, generalized circuits may contain *cycles*. Third, the constraints on the nodes are *approximate* depending on a precision parameter $\varepsilon$. This enables generalized circuits to express a large class of approximate-fixed-point problems. An $\varepsilon$-*solution* to a generalized circuit is an assignment of values in $[0,1]$ to the nodes consistent with the constraints induced by the gates for precision $\varepsilon$. While an $\varepsilon$-solution always exists, the search problem $\varepsilon$-GCircuit of finding such an $\varepsilon$-solution is PPAD-complete for a sufficiently small constant $\varepsilon$ (Rubinstein, 2018).

In this paper, we reveal a conceptual flaw in the definition of the generalized circuit concept, namely that the solution concept is not *monotonic*. By this we mean that if $\varepsilon < \varepsilon'$, then an $\varepsilon$-solution to a given generalized circuit is not necessarily also an $\varepsilon'$-solution to the same circuit. The issue lies with the Boolean gates NOT, AND, and OR and the way how these gates operate on *approximately Boolean values* (Section 3).

Not having monotonicity violates our intuition for an approximate solution concept. For example, the simple idea that finding an $\varepsilon$-approximate solution gets (weakly) harder as $\varepsilon$ gets smaller implicitly relies on the assumption of monotonicity.

To overcome this problem of non-monotonicity, we introduce a new variant of the generalized circuits problem that has stronger constraints for the Boolean gates that satisfy monotonicity. We call this variant $\varepsilon$-GCircuit$^{\text{SB}}$ ("SB" for "stronger Boolean").[1] $\varepsilon$-GCircuit$^{\text{SB}}$ serves as a monotonic drop-in replacement for $\varepsilon$-GCircuit in hardness proofs about generalized circuits themselves. In a second step, we show that Boolean gates (our stronger version or the original weaker version) are in fact a redundant feature: we can represent each of the Boolean gates using only the other (arithmetic and comparison) gate types. Our result implies that two new monotonic search problems are PPAD-complete: $\varepsilon$-GCircuit$^{\text{SB}}$ and the restriction $\varepsilon$-GCircuit$^{\text{NB}}$ of $\varepsilon$-GCircuit where no Boolean gates are allowed ("NB" for "no Boolean"; see Section 4).

We then illustrate that the lack of monotonicity in $\varepsilon$-GCircuit has led to several subtle technical issues in prior work (to be precise, in Chen, Deng and Teng (2009) and Rubinstein (2018)) that, to the best of our knowledge, have been overlooked so far. While these issues are of a mere technical nature and can be circumvented using more careful argumentation, we demonstrate that $\varepsilon$-GCircuit$^{\text{SB}}$ can be used as a drop-in replacement for $\varepsilon$-GCircuit in these pieces of work and provides a clean and conceptually simple way to eliminate these issues (Section 5).

---

[1] The distinguishing feature of $\varepsilon$-GCircuit$^{\text{SB}}$, which makes the solution concept monotonic, is shared by a variant of generalized circuits considered earlier by Othman, Papadimitriou and Rubinstein (2016). However, this variant is heavily customized to their application (fair allocation), while we aim to stay as faithful to the standard definition of generalized circuits as possible. The authors also did not discuss the relevance of monotonicity.

We argue that, due to the desirability of monotonicity, the $\varepsilon$-GCircuit problem in its current form is difficult to work with and future studies of generalized circuits should either consider the $\varepsilon$-GCircuit$^{\text{SB}}$ problem or the $\varepsilon$-GCircuit$^{\text{NB}}$ problem (i.e., leave out Boolean gates altogether). Monotonic solution concepts match our expectations and are thus easier to reason about. The fact that the Boolean gates are optional will simplify reductions from generalized circuits to other problems. We hope that this will enable new complexity results for equilibrium search problems in the future.

## 2 Preliminaries: Generalized Circuits

We follow the definition of a generalized circuit in Rubinstein (2018). A *generalized circuit* is a collection of *nodes* and *gates*, where each node is labeled as an *input* of any number of gates (including zero) and as an *output* of at most one gate. Inputs to the same gate are distinguishable from each other. Each gate has one of the types $C_\zeta$, $C_{\times\zeta}$, $C_=$, $C_+$, $C_-$, $C_<$, $C_\vee$, $C_\wedge$, or $C_\neg$. For the gate types $C_\zeta$ and $C_{\times\zeta}$, a numeric parameter $\zeta$ is specified in addition to their input and output nodes. The *length* of a generalized circuit is the number of bits needed to describe the circuit, including the nodes, the mapping from nodes to inputs and outputs of gates, and numeric parameters $\zeta$ involved.

For any $\varepsilon > 0$, an $\varepsilon$-*approximate solution* (or $\varepsilon$-*solution* for short) of a generalized circuit is a mapping $x$ that assigns to each node $v$ a value $x[v] \in [0, 1]$ such that at each gate, the constraints in Table 1 hold. We write $[x] := \min(1, \max(0, x))$ and we write $y = x \pm \varepsilon$ to mean that $|x - y| \leq \varepsilon$. $\varepsilon$-GCircuit is the search problem of finding an $\varepsilon$-solution of a given generalized circuit. It is easy to show that an $\varepsilon$-solution always exists (using Kakutani's fixed-point theorem), has polynomial length, and that $\varepsilon$-GCircuit is in PPAD. This is true even if $\varepsilon$ decreases exponentially with the input size.

The gates can be grouped into three categories: the *arithmetic gates* $C_\zeta$, $C_{\times\zeta}$, $C_=$, $C_+$, and $C_-$, the *comparison gate* $C_<$, and the *Boolean gates* $C_\vee$, $C_\wedge$, and $C_\neg$. Note from Table 1 how the comparison gate is *brittle*: its output value is unconstrained in $[0, 1]$ if $x[a_1]$ and $x[a_2]$ are $\varepsilon$-close to each other. This is crucial to guarantee existence of an $\varepsilon$-solution[2] and it is also necessary to enable reductions from generalized circuits to other approximate solution concepts like approximate Nash equilibrium, where an exact comparison gadget may not be attainable (see Daskalakis, Goldberg and Papadimitriou (2009) for a discussion). The Boolean gates are defined in a similar way, operating on *approximately Boolean values*. That is, we consider any value within $[0, \varepsilon]$ Boolean FALSE and any value within $[1-\varepsilon, 1]$ Boolean TRUE. The Boolean gates are then only required to return an approximately Boolean value at their output if their inputs are also approximately Boolean. If the inputs are not approximately Boolean, i.e., if they lie in the interval $(\varepsilon, 1-\varepsilon)$, any output value is allowed. For example, the $C_\neg$ gate can map an input 0.5 to any number in $[0, 1]$. This provides a minimal specification of "approximate Boolean gates" and is important for reductions because the problem one wants to reduce to may only be able to express Boolean functions up to such errors (e.g., approximate fixed point problems, see Section 3 below). Note further how the arithmetic gates accumulate errors (a chain of, say $n$ $C_=$ gates has a total error of $n\varepsilon$) while the Boolean gates do not. This is exploited, for example, in Rubinstein (2018).

---
[2]A brittle comparison gadget, and thus all gadgets, can be represented by a continuous function with Lipschitz constant $O(1/\varepsilon)$. By Brouwer's fixed point theorem and rounding, an $\varepsilon$-approximate solution of length $O(1/\text{Length}(\varepsilon)^2)$ always exists.

**Table 1** Conditions required to hold at a gate of the respective type with input nodes $a_i$ and output node $v$ in an $\varepsilon$-solution for a generalized circuit. For each gate, one output node and between 0 and 2 input nodes (depending on the gate type) are specified. For the gates $C_\zeta$ and $C_{\times\zeta}$, an additional numeric parameter $\zeta \in [0,1]$ is specified.

| Gate Type | Short | Constraint |
|---|---|---|
| Constant | $C_\zeta$ | $x[v] = \zeta \pm \varepsilon$ |
| Scaling | $C_{\times\zeta}$ | $x[v] = [\zeta \cdot x[a_1]] \pm \varepsilon$ |
| Copy | $C_=$ | $x[v] = x[a_1] \pm \varepsilon$ |
| Addition | $C_+$ | $x[v] = [x[a_1] + x[a_2]] \pm \varepsilon$ |
| Subtraction | $C_-$ | $x[v] = [x[a_1] - x[a_2]] \pm \varepsilon$ |
| Comparison | $C_<$ | $x[a_1] < x[a_2] - \varepsilon \Rightarrow x[v] \leq \varepsilon$ <br> $x[a_1] > x[a_2] + \varepsilon \Rightarrow x[v] \geq 1 - \varepsilon$ |
| OR | $C_\vee$ | $x[a_1] \leq \varepsilon$ and $x[a_2] \leq \varepsilon \Rightarrow x[v] \leq \varepsilon$ <br> $x[a_1] \geq 1-\varepsilon$ or $x[a_2] \geq 1-\varepsilon \Rightarrow x[v] \geq 1-\varepsilon$ |
| AND | $C_\wedge$ | $x[a_1] \leq \varepsilon$ or $x[a_2] \leq \varepsilon \Rightarrow x[v] \leq \varepsilon$ <br> $x[a_1] \geq 1-\varepsilon$ and $x[a_2] \geq 1-\varepsilon \Rightarrow x[v] \geq 1-\varepsilon$ |
| NOT | $C_\neg$ | $x[a_1] \leq \varepsilon \Rightarrow x[v] \geq 1-\varepsilon$ <br> $x[a_1] \geq 1-\varepsilon \Rightarrow x[v] \leq \varepsilon$ |

$\varepsilon$-GCIRCUIT is known to be PPAD-complete for a sufficiently small constant $\varepsilon$ (Rubinstein, 2018). Thus, no polynomial-time approximation scheme (PTAS) exists unless P=PPAD. This is the strongest hardness result for $\varepsilon$-GCIRCUIT known to date.

## 3  $\varepsilon$-GCIRCUIT Does Not Satisfy Monotonicity

We now formally define monotonicity and we show that $\varepsilon$-GCIRCUIT does not in general satisfy it. Let $X$ be a set and let $P_\varepsilon : X \to \{\text{TRUE}, \text{FALSE}\}$ for $0 < \varepsilon < 1$ be a *family of properties* of elements of $X$. We call the family $P$ *monotonic* if for all $\varepsilon < \varepsilon'$ and all $x \in X$, $P_\varepsilon(x)$ implies $P_{\varepsilon'}(x)$.

Essentially anything we would call an "approximate solution concept" is monotonic. For example, if $G$ is a game, $X$ is the set of mixed strategy profiles of $G$, and $P_\varepsilon(x) = \text{TRUE}$ iff $x$ is an $\varepsilon$-approximate Nash equilibrium of $G$, then the family $P$ is monotonic. Likewise, well-supported approximate Nash equilibria (Daskalakis, Goldberg and Papadimitriou, 2009) and relative approximate Nash equilibria are monotonic. Another important family of monotonic properties are *approximate fixed points*. Let $n \geq 1$, $X = [0, 1]^n$, and let $F : X \to X$ be a function. Let $P_\varepsilon(x) = \text{TRUE}$ iff $F_i(x) = x_i \pm \varepsilon$ for all $i$. $x$ is then called an *$\varepsilon$-approximate fixed point* of $F$. $P$ is obviously monotonic.[3]

We now show that "$\varepsilon$-solution to a certain generalized circuit" is *not* in general monotonic.

**Proposition 1.** *There exists a generalized circuit $C$ such that the family of properties $P_\varepsilon(x) := $ "$x$ is an $\varepsilon$-solution for $C$" is not monotonic.*

*Proof.* Let $C$ consist of two nodes $a$ and $v$ connected by a single $C_\neg$ gate. Let $0 < \varepsilon < 1/4$ and let $x[a] = 1.5\varepsilon$ and $x[v] = 0.5$. Since $x[a] \notin [0; \varepsilon] \cup [1 - \varepsilon; 1]$, the $C_\neg$ gate does not constrain the value of the output and thus $x$ is an $\varepsilon$-solution. Let $\varepsilon' = 2\varepsilon$. Now $x[a] \leq \varepsilon'$, so the $C_\neg$ gate requires that $x[v] \geq 1 - \varepsilon'$. But this is not the case. Thus, $x$ is not an $\varepsilon'$-solution, which violates monotonicity. □

*Remark* 1. In the specific, stylized example in the above proof, there are of course many $\varepsilon$-solutions that are also $\varepsilon'$-solutions for $\varepsilon' > \varepsilon$, like $(x[a] = 0, x[v] = 1)$. One might argue that one should only consider these "normal" solutions and that the $\varepsilon$-solution we discuss is pathological. If the gate is part of a larger circuit, however, it is not clear anymore how one would transition to a "normal" solution while still satisfying the constraints at all gates. We discuss this in Appendix B.

At a conceptual level, the reason why "$\varepsilon$-solution to a generalized circuit" is not monotonic is because for the Boolean gates, the respective constraint is a collection of implications where conditions like $x \leq \varepsilon$ and $x \geq 1 - \varepsilon$ occur on both sides of each implication (see Table 1). Both sides get weaker as $\varepsilon$ is increased and thus the effect on the overall constraint is ambiguous. Indeed, it is not hard to construct analogous counterexamples to the proof of Proposition 1 for the $C_\vee$ and $C_\wedge$ gates. Note that, in contrast to the Boolean gates, the comparison gate is not affected by this problem. This is because here, the left-hand side of the implication becomes *stronger* as $\varepsilon$ is increased, so the whole implication becomes unambiguously weaker.

The fact that "$\varepsilon$-solution to a generalized circuit" is not monotonic violates our intuition for approximation problems. For example, we would typically assume that the $\varepsilon$-GCircuit search problem becomes (weakly) harder when we decrease $\varepsilon$. This is of course based on the assumption that a solution to $\varepsilon$-GCircuit will also be a solution to $\varepsilon'$-GCircuit if $\varepsilon < \varepsilon'$ (i.e., monotonicity). However, since monotonicity is not guaranteed, we cannot immediately exclude the possibility that the problem becomes *easier* again when we decrease $\varepsilon$ far enough. This might be because for a low $\varepsilon$, many inputs to Boolean gates can be chosen to be not approximately Boolean and so the outputs of these gates can be arbitrary, giving us additional degrees of freedom to satisfy other constraints. We show in Appendix B that the problem does not actually become easier in a computational complexity sense, but this requires careful argumentation.

---
[3]Approximate fixed points occur in many search problems, where $F$ is then defined in some way based on the input. A related concept are *strong approximate fixed points* (Etessami and Yannakakis, 2010), which need to be close to an exact fixed point.

## 4 Restoring Monotonicity

We have just seen that the lack of monotonicity creates subtle pitfalls in otherwise trivial arguments. In fact, we will demonstrate in Section 5 that non-monotonicity can lead to many more issues, including in prior work. To overcome this problem, we now present a way to restore monotonicity. We will show in Section 5 that our approach eliminates the above-mentioned issues at a conceptual level.

Non-monotonicity arises due to the Boolean gates. Of course, we cannot simply remove the Boolean gates from consideration because the hardness proofs for $\varepsilon$-GCircuit rely on having access to Boolean gates. Instead, to restore monotonicity, we define a new variant of the problem that has stronger constraints for the Boolean gates that satisfy monotonicity. We call this variant $\varepsilon$-GCircuit$^{\text{SB}}$ ("SB" for "stronger Boolean"). $\varepsilon$-GCircuit$^{\text{SB}}$ is very useful for hardness proofs about generalized circuits themselves (see Section 5). However, the fact that we have strengthened the Boolean gates creates two new problems: first, reductions from $\varepsilon$-GCircuit to other problems do not automatically provide reductions for the stronger Boolean gates. Second, it is not clear at this point that $\varepsilon$-GCircuit$^{\text{SB}}$ is in PPAD. To resolve *these* problems, we show that Boolean gates (our stronger version or the original weaker version) are a redundant feature: we can represent each of the Boolean gates using only the other (arithmetic and comparison) gate types.[4] This immediately implies that the restriction of $\varepsilon$-GCircuit where no Boolean gates are allowed, and which we call $\varepsilon$-GCircuit$^{\text{NB}}$ ("NB" for "no Boolean"), is already PPAD-complete. Note that $\varepsilon$-GCircuit$^{\text{NB}}$ is also monotonic.

**Table 2** Conditions required to hold at a gate $g$ with input nodes $a_i$ and output node $v$ in a strong $\varepsilon$-solution for a generalized circuit. The constraints differ from Table 1 only with regards to the Boolean gates (highlighted in gray).

| Gate Type | Short | Constraint |
|---|---|---|
| Constant | $C_\zeta$ | $x[v] = \zeta \pm \varepsilon$ |
| Scaling | $C_{\times\zeta}$ | $x[v] = [\zeta \cdot x[a_1]] \pm \varepsilon$ |
| Copy | $C_=$ | $x[v] = x[a_1] \pm \varepsilon$ |
| Addition | $C_+$ | $x[v] = [x[a_1] + x[a_2]] \pm \varepsilon$ |
| Subtraction | $C_-$ | $x[v] = [x[a_1] - x[a_2]] \pm \varepsilon$ |
| Comparison | $C_<$ | $x[a_1] < x[a_2] - \varepsilon \Rightarrow x[v] \leq \varepsilon$ <br> $x[a_1] > x[a_2] + \varepsilon \Rightarrow x[v] \geq 1 - \varepsilon$ |
| OR | $C_\vee$ | $x[a_1] < 1/2 - \varepsilon$ and $x[a_2] < 1/2 - \varepsilon \Rightarrow x[v] \leq \varepsilon$ <br> $x[a_1] > 1/2 + \varepsilon$ or $\ \ x[a_2] > 1/2 + \varepsilon \Rightarrow x[v] \geq 1 - \varepsilon$ |
| AND | $C_\wedge$ | $x[a_1] < 1/2 - \varepsilon$ or $\ \ x[a_2] < 1/2 - \varepsilon \Rightarrow x[v] \leq \varepsilon$ <br> $x[a_1] > 1/2 + \varepsilon$ and $x[a_2] > 1/2 + \varepsilon \Rightarrow x[v] \geq 1 - \varepsilon$ |
| NOT | $C_\neg$ | $x[a_1] < 1/2 - \varepsilon \Rightarrow x[v] \geq 1 - \varepsilon$ <br> $x[a_1] > 1/2 + \varepsilon \Rightarrow x[v] \leq \varepsilon$ |

## 4.1 The $\varepsilon$-GCIRCUIT$^{\text{SB}}$ Problem

Recall that non-monotonicity of the $\varepsilon$-solution concept arises because expressions of the form $x[a_1] \geq 1 - \varepsilon$ (which occur on the left-hand side of the constraints for Boolean gates) become *weaker* as $\varepsilon$ increases. To resolve this, we replace these conditions so that they become *stronger* as $\varepsilon$ increases, so that the whole implication becomes weaker. More in detail, we replace expressions of the form $x[a_1] \geq 1 - \varepsilon$ by expressions of the form $x[a_1] > 1/2 + \varepsilon$ on the left-hand side of the implication for Boolean gates.[5] This yields the constraints in Table 2. We call an assignment $x$ that satisfies these constraints a *strong $\varepsilon$-solution* and we call the corresponding search problem $\varepsilon$-GCIRCUIT$^{\text{SB}}$.[6] This restores monotonicity.

**Proposition 2.** *Let $C$ be a generalized circuit.*

---

[4]Of course, many more gates beyond the Boolean gates are redundant or could be replaced by simplified versions of the respective gate. For example, $C_=$ can be replaced by $C_{\times 1}$. However, only the Boolean gates are relevant for monotonicity. See Othman, Papadimitriou and Rubinstein (2016) and Schuldenzucker, Seuken and Battiston (2019) for reduced sets of gates.

[5]Note that we also replace weak by strict inequalities in the process. We do this to simplify the following arguments in this paper and to receive the *continuity* property discussed in Remark 2 below. It is not crucial for our construction, though.

[6]A similar variant of the generalized circuits problem was first studied by Othman, Papadimitriou and Rubinstein (2016). The authors introduced an additional parameter $\beta = \Theta(\varepsilon)$ and then specified the Boolean gates like in Table 2 if we replace $\varepsilon$ by $\beta$ on the left-hand sides of the Boolean gates. Their variant differs from the variant we describe here in other details of the problem. For example, scaling is only allowed by a factor $1/2$ and the definition of the $C_\neg$ gate is not analogous to the two other Boolean gates. In this paper, we aim for the smallest deviation from Rubinstein's (2018) variant that eliminates the aforementioned problems.

1. *For any $\varepsilon < 1/4$, any strong $\varepsilon$-solution of $C$ is also an $\varepsilon$-solution of $C$*

2. *The family of properties $P_\varepsilon(x) :=$ "$x$ is a strong $\varepsilon$-solution of $C$" is monotonic.*

*Proof.* 1: We can consider each gate separately and we only need to consider the Boolean gates, since the constraints for the other gates are the same between $\varepsilon$-GCIRCUIT and $\varepsilon$-GCIRCUIT$^{\text{SB}}$. For the Boolean gates, note that for all $z \in [0,1]$ and $\varepsilon < 1/4$ we have $z \leq \varepsilon \Rightarrow z < 1/2 - \varepsilon$ and $z \geq 1 - \varepsilon \Rightarrow z > 1/2 + \varepsilon$. Thus, whenever we require $x[v] = 0 \pm \varepsilon$ or $x[v] = 1 \pm \varepsilon$ in an $\varepsilon$-solution, we make the same requirement in a strong $\varepsilon$-solution. Therefore, every strong $\varepsilon$-solution satisfies the constraints for an $\varepsilon$-solution.

2: Again, we only need to consider the constraints corresponding to Boolean gates since we have already seen that the others satisfy monotonicity. For the Boolean gates, consider the $C_\neg$ gate, let $x$ be an $\varepsilon$-solution and let $\varepsilon < \varepsilon'$. We distinguish the two cases in the constraint for the $C_\neg$ gate for a strong $\varepsilon'$-solution.

- If $x[a_1] < 1/2 - \varepsilon'$, then $x[a_1] < 1/2 - \varepsilon$ and thus, since $x$ is an $\varepsilon$-solution, $x[v] = 1 \pm \varepsilon = 1 \pm \varepsilon'$ as required.

- If $x[a_1] > 1/2 + \varepsilon'$, then likewise $x[a_1] > 1/2 + \varepsilon$ and thus $x[v] = 0 \pm \varepsilon = 0 \pm \varepsilon'$.

The proofs for the other two Boolean gates are analogous. □

Due to monotonicity, the $\varepsilon$ parameter of $\varepsilon$-GCIRCUIT$^{\text{SB}}$ now behaves as we would intuitively expect. For example, the $\varepsilon$-GCIRCUIT$^{\text{SB}}$ problem trivially becomes (weakly) harder as $\varepsilon$ decreases.

*Remark* 2 (Continuity of the solution concept). $\varepsilon$-GCIRCUIT$^{\text{SB}}$ is distinguished from $\varepsilon$-GCIRCUIT by another intuitive property that we call *continuity* of the solution concept.[7] By *continuity* we mean the following. Fix a generalized circuit and let $x^n \to x$ and $\varepsilon^n \to \varepsilon$ be two convergent sequences such that $x^n$ is a strong $\varepsilon^n$-solution for all $n$; then $x$ is a strong $\varepsilon$-solution. In particular, if $\varepsilon^n \to 0$, then $x$ is a strong exact solution. Continuity holds for $\varepsilon$-GCIRCUIT$^{\text{SB}}$ because only *strict* inequalities appear on the left-hand sides of the constraints for the Boolean gates.[8] In $\varepsilon$-GCIRCUIT, these inequalities are weak and $\varepsilon$-GCIRCUIT does not satisfy continuity.[9] Note that continuity of the solution concept is orthogonal to monotonicity and does not affect any of the other results, and in particular it does not affect hardness of the problem.

---

[7] We thank Xi Chen for bringing this property to our attention.

[8] More in detail, note that continuity for $\varepsilon$-GCIRCUIT$^{\text{SB}}$ is equivalent to closedness of the set $S := \{(x, \varepsilon) \mid x \text{ is a strong } \varepsilon\text{-solution for } C\}$ for any generalized circuit $C$. This in turn holds because $S = \bigcap_{g \text{ gate}} \{(x, \varepsilon) \mid x \text{ satisfies the constraint in Table 2 for } g \text{ with tolerance } \varepsilon\}$ and all of these sets are closed. The first constraint of the $C_\neg$ gate, for example, is equivalent to $x[a_1] \geq 1/2 - \varepsilon \lor x[v] \geq 1 - \varepsilon$, which obviously leads to a closed set. Also recall that we write $y = x \pm \varepsilon$ for the *weak* inequalities $x - \varepsilon \leq y \leq x + \varepsilon$. This is crucial for continuity at the arithmetic gates and differs from, e.g., Rubinstein (2018).

[9] To see that $\varepsilon$-GCIRCUIT is not continuous, consider a single $C_\neg$ gate, let $\varepsilon^n = 1/n$, $x^n[a_1] = 2/n$, and $x^n[v] = 0$. WLOG assume $n \geq 4$. Note that $x^n[a_1] \not\leq \varepsilon^n$ and $x^n[a_1] \not\geq 1 - \varepsilon^n$, so $x^n$ is an $\varepsilon^n$-solution for all $n$. However, $x[a_1] = 0 \leq 0 = \varepsilon$, but $x[v] = 0 \not\geq 1 = 1 - \varepsilon$, so $x$ is not an $\varepsilon$-solution.

Note further that continuity does *not* imply any statement regarding the "speed of convergence." That is, we do not receive an upper bound on $\|x - x^n\|$ dependent on $|\varepsilon - \varepsilon^n|$. In light of the hardness results in Etessami and Yannakakis (2010) regarding *strong fixed points*, such a result seems unlikely to be obtainable.

$\varepsilon$-GCircuit$^{SB}$ is monotonic and offers access to Boolean gates, which makes it useful to study the hardness of generalized circuit problems themselves. However, since we made the Boolean gates stronger, it might be the case that $\varepsilon$-GCircuit$^{SB}$ is a strictly harder problem than $\varepsilon$-GCircuit. It is an immediate consequence of the discussion in the following section that this is not the case. It will turn out that (even our stronger) Boolean gates do not actually add any expressiveness on top of the other gates.

## 4.2 Redundancy of Boolean Gates

Perhaps surprisingly, we can construct the Boolean gates in the definition of $\varepsilon$-GCircuit$^{SB}$ from the arithmetic and comparison gates. Thus, the Boolean gates are redundant as a feature and we receive a reduction from $\Theta(\varepsilon)$-GCircuit$^{NB}$ to $\varepsilon$-GCircuit$^{SB}$. Recall that $\varepsilon$-GCircuit$^{NB}$ is the restriction of $\varepsilon$-GCircuit where no Boolean gates are allowed. In the following, we write "$\varepsilon \ll 1$" (read: "$\varepsilon$ sufficiently small") to mean that a statement holds for all $\varepsilon$ below a certain positive threshold. Unless indicated, the threshold is a constant that does not depend on the context of the statement.

**Lemma 1.** *For any generalized circuit $C$ we can construct in polynomial time a circuit $C'$ such that i) the nodes of $C'$ are a superset of the nodes of $C$, ii) $C'$ does not contain any Boolean gates, and iii) for any $\varepsilon \ll 1$, any $\varepsilon/2$-solution for $C'$ induces a strong $\varepsilon$-solution for $C$ via restriction to the nodes of $C$.*

*Proof.* We need to model the Boolean gates. The $C_\wedge$ gate is redundant because it can be expressed using $C_\neg$ and $C_\vee$ (recall that Boolean gates do not accumulate errors!). Assume $\varepsilon < 1/3$, let $\varepsilon' = \varepsilon/2$, and consider an $\varepsilon'$-solution.

We model $C_\neg$ as the expression $a_1 < 1/2$ using a $C_>$ and a $C_\zeta$ gate. Call the output of the $C_\zeta$ gate $z$. We have $x[z] = 1/2 \pm \varepsilon'$. If $x[a_1] < 1/2 - \varepsilon$, then $x[a_1] < x[z] - \varepsilon'$ and thus the output of the comparison gate is $1 \pm \varepsilon' = 1 \pm \varepsilon$. Likewise for $x[a_1] > 1/2 + \varepsilon$.

We model $C_\vee$ as $(a_1 > 1/2) + (a_2 > 1/2) > 1/2$. If one of $x[a_1]$ or $x[a_2]$ is $> 1/2 + \varepsilon$, then like above, the respective inner $C_>$ gate will return $1 \pm \varepsilon'$ and thus the output of the $C_+$ gate is $1 \pm 2\varepsilon' > 1/2 + \varepsilon'$ and the outer $C_>$ gate returns $1 \pm \varepsilon' = 1 \pm \varepsilon$. If $x[a_1], x[a_2] < 1/2 - \varepsilon$, then both inner $C_>$ gates return $0 \pm \varepsilon'$, $C_+$ returns a value $\leq 2\varepsilon' < 1/2 - \varepsilon'$, and the final $C_>$ gate returns $0 \pm \varepsilon' = 0 \pm \varepsilon$. □

Note that Daskalakis, Goldberg and Papadimitriou (2009) previously argued that one might try to simulate Boolean gates using arithmetic gates, expressing $x \vee y$ as $[x + y]$ and $\neg x$ as $1 - x$. However, they also noted that this would lead to an accumulation of errors when several of these gates are put in a row. One would thus not be able to satisfy the constraints for the Boolean gates (in a strong or a weak $\varepsilon$-solution) this way. Daskalakis, Goldberg and Papadimitriou then define dedicated Boolean game gadgets that do not accumulate errors. In Lemma 1, we have shown that such dedicated gadgets (or *gates* in the abstract problem description) are not actually required: we can represent the Boolean gates using the arithmetic gates if we in addition employ comparison gates to prevent error accumulation.[10]

The lemma immediately implies:

**Corollary 1.** *The problems $\varepsilon$-GCircuit$^{SB}$, $\varepsilon$-GCircuit, and $\varepsilon$-GCircuit$^{NB}$ are all PPAD-complete for $\varepsilon \ll 1$:*

*Proof.* We have:

$$\varepsilon/2\text{-GCircuit}^{NB} \geq_P \varepsilon\text{-GCircuit}^{SB} \geq_P \varepsilon\text{-GCircuit} \geq_P \varepsilon\text{-GCircuit}^{NB}$$

where "$\geq_P$" stands for polynomial-time reducibility. The first relation is by Lemma 1 and the others are trivial.

It is well-known that $\varepsilon$-GCircuit is in PPAD for all $\varepsilon > 0$. Thus, all of the problems are in PPAD for all $\varepsilon > 0$. For PPAD-hardness for $\varepsilon \ll 1$, it is enough to show that $\varepsilon$-GCircuit$^{SB}$ or $\varepsilon$-GCircuit are PPAD-hard for $\varepsilon \ll 1$. This follows via Rubinstein's (2018) hardness proof for $\varepsilon$-GCircuit. We defer a discussion of this proof to Section 5, where we show that the proof is not affected by an implicit monotonicity assumption and further applies to $\varepsilon$-GCircuit$^{SB}$ without modification. □

Corollary 1 is useful because it implies that, when performing reductions from generalized circuits, there is no need to provide a reduction for the Boolean gates. In particular, via Lemma 1, all reductions from $\varepsilon$-GCircuit to other problems in prior work also provide a reduction from $\varepsilon$-GCircuit$^{SB}$.

## 5 Eliminating Issues With Non-Monotonicity in Prior Work

To the best of our knowledge, monotonicity has not been discussed in any piece of prior work on generalized circuits. This raises the question whether or not it has been carefully considered in the past. As explained in the previous section, mere reductions *from* $\varepsilon$-GCircuit to other problems automatically provide reductions from $\varepsilon$-GCircuit$^{SB}$ and will therefore not be affected. We thus take a close look at those pieces of work where hardness of the $\varepsilon$-GCircuit problem itself and its variants is established. Specifically, we discuss the three foundational papers on generalized circuits: Daskalakis, Goldberg and Papadimitriou (2009), Chen, Deng and Teng (2009), and Rubinstein (2018). We show that the first of these papers is not affected by non-monotonicity while in contrast, non-monotonicity does create subtle technical issues in the latter two. We then show how replacing $\varepsilon$-GCircuit by $\varepsilon$-GCircuit$^{SB}$ eliminates these issues. $\varepsilon$-GCircuit$^{SB}$ serves as a drop-in replacement for $\varepsilon$-GCircuit, allowing us to keep all unaffected arguments the same.

---

[10]It should be noted that the Boolean game gadgets in Daskalakis, Goldberg and Papadimitriou (2009) *do* satisfy a monotonic definition of the Boolean gates that is of intermediate strength between $\varepsilon$-GCircuit and $\varepsilon$-GCircuit$^{SB}$. For the OR game gadget, for example, we have that $x[a_1] + x[a_2] > 1/2 + \varepsilon \Rightarrow x[v] = 1$ and $x[a_1] + x[a_2] < 1/2 - \varepsilon \Rightarrow x[v] = 0$. This is not quite enough for a strong $\varepsilon$-solution, but it is a monotonic property by itself. We discuss another such intermediate definition of Boolean gates in Appendix C.

We would like to stress that the purpose of our discussion is not to diminish the contributions of these seminal papers. Rather, we find it instructive to demonstrate what problems non-monotonicity can create by using the proofs in the three seminal papers as examples, rather than inventing examples ourselves. Note that the issues in prior work that we point out are of a mere technical nature and could be circumvented by careful argumentation. We present a way how this could be done without relying on $\varepsilon$-GCircuit$^{SB}$ in Appendix B. However, as we will see, $\varepsilon$-GCircuit$^{SB}$ provides a particularly clean and conceptually simple solution to these problems. We will now go through the proof steps in the three papers one by one. We will label the issues #1–#5, to refer back to them in the appendix.

## 5.1 Daskalakis, Goldberg and Papadimitriou (2009)

Daskalakis, Goldberg and Papadimitriou (2009) were the first to prove PPAD-hardness for the problem of finding an approximate Nash equilibrium, for an exponentially small $\varepsilon$. The proof is by reduction from a variant of the Brouwer fixed-point problem using a collection of *game gadgets*. These game gadgets correspond to a variant of $\varepsilon$-GCircuit where the Boolean gates are defined via *exact* rather than approximately Boolean values. For example, the output of the NOT gate is 1 if the input is 0, 0 if the input is 1, and unrestricted otherwise. The comparison gate also yields an exact Boolean value rather than an approximately Boolean one. In contrast to the (nowadays more standard) definition of generalized circuits we have presented in Section 2, their variant of $\varepsilon$-GCircuit satisfies monotonicity. Thus, this paper is not affected.

## 5.2 Chen, Deng and Teng (2009)

Chen, Deng and Teng (2009) proved PPAD-hardness of finding an approximate Nash equilibrium in a two-player game for polynomially small $\varepsilon$. En-route to this result, the authors provide the first explicit definition of the generalized circuit concept. In this early variant, values of nodes are truncated to $[0, 1/K]$ rather than $[0, 1]$, where $K$ is the number of nodes of the circuit. Note that $\varepsilon$ has to decrease at least linearly in $K$, otherwise the error term $\varepsilon$ would eventually become larger than the range of the solution and the problem would become trivial. We call this variant of the problem $\varepsilon$-GCIRCUIT$_\text{C}$ to distinguish it from Rubinstein's, nowadays more standard, variant. It is easy to see that $\varepsilon$-GCIRCUIT is computationally equivalent to $\varepsilon/K$-GCIRCUIT$_\text{C}$ via scaling. Like Chen, Deng and Teng, we write POLY$^c$-GCIRCUIT$_\text{C}$ for $K^{-c}$-GCIRCUIT$_\text{C}$.

The hardness proof in Chen, Deng and Teng (2009) proceeds in three steps. (1) The authors establish hardness of a variant of the BROUWER approximate fixed-point problem. (2) They reduce this problem to POLY$^3$-GCIRCUIT$_\text{C}$. (3) They reduce POLY$^3$-GCIRCUIT$_\text{C}$ to the problem of finding an $n^{-12}$-approximate Nash equilibrium in a two-player game, where $n$ is the number of actions. The last step is carried out using a collection of two-player game gadgets. Two additional reductions establish that the exponents do not actually matter for the complexity of the problems.

A part that demands some scrutiny is step 2, where Brouwer is reduced to POLY$^3$-GCIRCUIT$_\text{C}$. Fortunately, detailed examination of the proof shows that no implicit monotonicity assumption is made. This is because a single $\varepsilon$ (namely *exactly* $\varepsilon = K^{-3}$) is considered over the whole course of the proof. The same is true for the description of the game gadgets (step 3).

A place that *does* suffer from an implicit monotonicity assumption is the "padding theorem" (Chen, Deng and Teng, 2009, Theorem 5.7), where the authors show that the hardness of the POLY$^c$-GCIRCUIT$_\text{C}$ problem does not increase if we increase $c$, as long as $c \geq 3$. The proof is by reduction from POLY$^{2b+1}$-GCIRCUIT$_\text{C}$ to POLY$^3$-GCIRCUIT$_\text{C}$, for any integer $b > 1$. However, since we do not have monotonicity, this only implies the statement for *odd integer* values of $c$. POLY$^4$-GCIRCUIT$_\text{C}$, for example, might still be a harder problem. Further, and again due to the lack of monotonicity, we only receive a statement for the $\varepsilon$-GCIRCUIT$_\text{C}$ problem where $\varepsilon$ is exactly of form $\varepsilon = n^{-c}$ for some $c$. We do not receive any statement for arbitrary polynomials like $2n^{-3} + n^{-2}$. We call this issue #1.

To resolve this issue, we can define an $\varepsilon$-GCIRCUIT$^\text{SB}$ analog to Chen, Deng and Teng's (2009) version of generalized circuits. To do this, we replace in Table 2 truncation to $[0, 1]$ by truncation to $[0, 1/K]$ and for the Boolean gates, we replace the constant $1/2$ by $1/(2K)$. We then consider the problem $K^{-c}$-GCIRCUIT$'_\text{C}$ where $c \geq 1$ is a constant. Note that, like before, $\varepsilon$ has to decrease in $K$ at least linearly. When applied to this variant of the generalized circuit concept, the proof in the paper yields:

**Proposition 3** (Chen, Deng and Teng (2009), Theorem 5.7 for $\varepsilon$-GCIRCUIT$^\text{SB}$). *For any $c \geq 3$, $K^{-c}$-GCIRCUIT$'_\text{C} \leq_\text{P} K^{-3}$-GCIRCUIT$'_\text{C}$.*

*Proof.* Since we now have monotonicity, it is enough to show the statement for every $c$ of form $c = 2b + 1$ where $b > 0$ is an integer. To this end, let $C$ be a generalized circuit with $K \geq 2$ nodes and let $\varepsilon = K^{-c}$. The proof of Theorem 5.7 in Chen, Deng and Teng (2009) constructs a new circuit with $K' := K^b = 1/K \cdot K^{1-b}$ nodes such that for $\varepsilon' := \varepsilon K^{1-b}$, any strong $\varepsilon'$-solution for the new circuit gives rise to a strong $\varepsilon$-solution for the original circuit via scaling by $K^{1-b}$. And $\varepsilon' = K'^{-3}$. □

## 5.3 Rubinstein (2018)

Rubinstein (2018) proved PPAD-hardness for the problem of finding an $\varepsilon$-approximate Nash equilibrium for a sufficiently small *constant* $\varepsilon$.[11] The proof proceeds in four steps. (1) The author establishes hardness of a new class of instances of the BROUWER problem with constant $\varepsilon$. (2) He reduces this problem to $\varepsilon$-GCIRCUIT for a certain constant $\varepsilon$. (3) The author shows, using an additional black-box reduction, that $\varepsilon$-GCIRCUIT is still hard for some $\varepsilon$ when we limit the fan-out[12] of each gate to 2. (4) The author employs the game gadgets from Daskalakis, Goldberg and Papadimitriou (2009) to reduce *this* problem to the problem of finding an approximate Nash equilibrium in a degree-3 graphical game. Based on the considerations at the start of this sub-section, we should now take a closer look at steps 2-4.

The main hardness proof for $\varepsilon$-GCIRCUIT (step 2) is a reduction from the problem of finding an $\varepsilon^{1/4}$-approximate fixed point of a certain function to $\varepsilon$-GCIRCUIT, for any sufficiently small $\varepsilon$, where the constructed $\varepsilon$-GCIRCUIT instance depends on $\varepsilon$. Detailed examination of the proof shows that none of the arguments implicitly assume monotonicity. As the construction can be performed for arbitrarily small $\varepsilon$, this indeed shows hardness of $\varepsilon$-GCIRCUIT for any sufficiently small $\varepsilon$ (and not just for *one specific* $\varepsilon$, which is not a priori clear when monotonicity is not given).

The lack of monotonicity *does* lead to several problems in step 3, a black-box reduction from any given generalized circuit to a circuit with fan-out 2 (Rubinstein, 2018, page 941). In this reduction, the outputs of each comparison or Boolean gate with a fan-out greater than 2 are distributed using binary trees of double negation gates. The outputs of arithmetic gates, in contrast, are first transformed into a unary bit representation, then the resulting Boolean values are distributed using the aforementioned trees of double negation gates, and finally each copy is converted back into its numeric form. This *distribution subroutine* has maximum fan-out 2 and guarantees that each of its outputs is equal to its input with an error of $\pm\varepsilon$ in any $\varepsilon^2$-solution. One thus has to reduce the allowed error to $\varepsilon' \in \Theta(\varepsilon^2)$.[13]

---

[11] To clarify the relationship between the three papers: the proof in Rubinstein (2018) is for the sub-class of polymatrix degree-3 graphical games, but does not extend to two-player games. It is therefore an unambiguous improvement upon the main result in Daskalakis, Goldberg and Papadimitriou (2009), who considered the same class of games and *exponentially small* $\varepsilon$, but not upon Chen, Deng and Teng (2009), who considered two-player games. In two-player games, the problem is likely not PPAD-hard for constant $\varepsilon$ (Rubinstein, 2018).

[12] The *fan-out* of a gate $g$ is the number of gates $g'$ such that the output node of $g$ is an input node of $g'$.

[13] Note that we have interchanged $\varepsilon'$ and $\varepsilon$ compared to Rubinstein's (2018) formulation of the theorem, notation-wise, to ensure consistency of notation within the present paper.

There are three problems with this reduction, all of which arise from non-monotonicity at Boolean gates and all of which can lead to a situation where some $\varepsilon'$-solution to the reduced circuit is not an $\varepsilon$-solution to the original circuit. We provide detailed examples for this in Appendix A. For the first issue, consider a Boolean gate that already has fan-out $\leq 2$. Since no changes are made to such gates and we do not have monotonicity, the $\varepsilon'$-solution to the reduced circuit may fail to be an $\varepsilon$-solution for the original circuit (call this issue #2). Next, there may be arithmetic gates with fan-out $> 2$ whose outputs feed into Boolean gates. Here, the distribution subroutine introduces an additional error, which may turn values from approximately-Boolean to not-approximately-Boolean and may thus not correctly copy these values (issue #3). Finally, there may be Boolean gates with fan-out $> 2$ that feed into arithmetic gates. For these gates, the fact that we use double negation gates to distribute the outputs creates additional degrees of freedom in the reduced circuit when the inputs to Boolean gates are not approximately Boolean and thus their outputs are arbitrary (issue #4).

When we replace $\varepsilon$-GCircuit by $\varepsilon$-GCircuit$^{\text{SB}}$, issue #2 is immediately resolved because $\varepsilon$-GCircuit$^{\text{SB}}$ has monotonicity. To eliminate issues #3 and #4, the fact that we use $\varepsilon$-GCircuit$^{\text{SB}}$ allows us to make a modification to Rubinstein's original proof to obtain the following lemma:

**Proposition 4** (Rubinstein's (2018) fan-out 2 reduction for $\varepsilon$-GCircuit$^{\text{SB}}$)**.** *For any $\varepsilon \ll 1$, there is an $\varepsilon' \in \Theta(\varepsilon^2)$ such that there is a polynomial-time reduction from $\varepsilon$-GCircuit$^{SB}$ to the restriction of $\varepsilon'$-GCircuit$^{SB}$ to maximum fan-out 2.*

*Proof.* Let $\bar{\varepsilon} = \varepsilon/3$. Assume that $\varepsilon \ll 1$ in a way to be made precise below. We perform the construction in Rubinstein (2018, Theorem 1.6) with respect to $\bar{\varepsilon}$ where we make the following modification: instead of differentiating between the outputs of Boolean/comparison vs. arithmetic gates, we *always* apply the distribution subroutine to the output of any gate with fan-out $> 2$. Recall that this subroutine has one input and any number of outputs and ensures that for a certain $\varepsilon' \in \Theta(\bar{\varepsilon}^2) = \Theta(\varepsilon^2)$, in an $\varepsilon'$-solution, each output equals the input up to an error of $\pm \bar{\varepsilon}$. It is easy to see that the distribution subroutine itself is not affected by any of the issues related to the fan-out 2 reduction. Assume that $\varepsilon' \leq \bar{\varepsilon}$.

Let $C$ be the original circuit, $C'$ the reduced circuit, and $x'$ a strong $\varepsilon'$-solution to $C'$. We show that the restriction of $x'$ to nodes in $C$ is a strong $\varepsilon$-solution for $C$. Let $g$ be a gate with inputs $a_1$ and $a_2$ (if any). Assume WLOG that distribution is applied to each of the inputs to $g$ and let $a_i'$ be the output of the respective distribution subroutine that is the new input to $g$ in $C'$. Let $v$ be the output of $g$ in $C$ and $C'$. We have $x'[a_i] = x'[a_i'] \pm \bar{\varepsilon}$ by the distribution subroutine. We perform case distinction over the type of $g$.

- If $g$ is an arithmetic gate, then for a sufficiently small $\varepsilon' \in \Theta(\varepsilon^2)$ it follows from

Lipschitz continuity that $x$ is an $\bar{\varepsilon}$-solution, and thus a strong $\varepsilon$-solution at $g$, just like in Rubinstein (2018).

- If $g$ is a comparison gate, assume WLOG that $x[a_1] < x[a_2] - \varepsilon$, i.e., $x[a_1] < x[a_2] - 3\bar{\varepsilon}$. By the distribution subroutine, $x'[a'_1] < x'[a'_2] - \bar{\varepsilon} \leq x'[a'_2] - \varepsilon'$ and thus, since $x'$ is a strong $\varepsilon'$-solution, $x'[v] \geq 1 - \varepsilon' \geq 1 - \varepsilon$ as required.

- If $g$ is a Boolean gate, consider any input $a_i$ to $g$. If $x'[a_i] < 1/2 - \varepsilon$, then $x'[a'_i] < 1/2 - \varepsilon + \bar{\varepsilon} \leq 1/2 - \varepsilon'$. Thus, if any input to $g$ is approximately Boolean FALSE w.r.t. $\varepsilon$ in $C$, then it is approximately Boolean FALSE w.r.t. $\varepsilon'$ in $C'$, and likewise for TRUE. Thus, if we require, based on the constraints, that $x'[v] \leq \varepsilon$ in a strong $\varepsilon$-solution of $C$, we require $x'[v] \leq \varepsilon'$ in a strong $\varepsilon'$-solution of $C'$. And the latter implies the former. Likewise for $x'[v] \geq 1 - \varepsilon$. □

Observe how the above proof eliminates issues #3 and #4 compared to Rubinstein's original proof. Issue #3 is eliminated in the last step (Boolean gates) and this crucially depends on the fact that we consider $\varepsilon$-GCIRCUIT$^{SB}$ instead of $\varepsilon$-GCIRCUIT: by monotonicity, we can choose $\varepsilon'$ sufficiently small to compensate for the additional error in the distribution subroutine. Issue #4 is eliminated because we use the distribution subroutine, which does not create additional degrees of freedom, for all gates.

The proposition together with Corollary 1 immediately yields:

**Corollary 2.** *For each of the problems $\varepsilon$-GCIRCUIT$^{SB}$, $\varepsilon$-GCIRCUIT, and $\varepsilon$-GCIRCUIT$^{NB}$, the restriction to maximum fan-out 2 is PPAD-complete for $\varepsilon \ll 1$.*

*Proof.* For $\varepsilon$-GCIRCUIT$^{SB}$, this follows from hardness of $\varepsilon$-GCIRCUIT$^{SB}$ and Proposition 4. For $\varepsilon$-GCIRCUIT$^{NB}$, we observe that the reduction in Lemma 1 preserves the fan-out 2 property. For $\varepsilon$-GCIRCUIT, it now follows trivially. □

Note that the lack of monotonicity introduces another subtle technical issue in Rubinstein (2018), specifically in the reduction from $\varepsilon$-GCIRCUIT to the problem of finding an approximate Nash equilibrium (step 4 in the outline of the proof above). Rubinstein uses the same game gadgets as Daskalakis, Goldberg and Papadimitriou (2009). However, for the Boolean game gadgets in that paper, we only know at this point that they work with exact Boolean values 0 and 1 in both the input and output, not necessarily approximately Boolean ones (see Section 5.1 above). And the former does not imply the latter because we do not have monotonicity. A priori, these gadgets might *rely* on receiving only values *exactly* 0 or 1 as their inputs (call this issue #5). Fortunately, this issue is eliminated immediately using Lemma 1: since the Boolean gates are redundant, it is not necessary to provide a reduction for them in the first place.[14]

---

[14]The way how we eliminated issue #5 may not feel satisfying to some readers because the

*Remark* 3 (Exact Boolean Gates). For Rubinstein (2018), there is another solution to the problems with non-monotonicity: rather than using $\varepsilon$-GCIRCUIT$^{\text{SB}}$, adopt the definition of generalized circuits from Daskalakis, Goldberg and Papadimitriou (2009), where only exact Boolean values are mapped to each other (see Section 5.1), and show hardness of this variant. For the case of graphical games, the game gadgets from Daskalakis, Goldberg and Papadimitriou (2009) provide a reduction from this variant with exact Boolean values. In many other applications, however, such a reduction is not possible. For example, any approximate fixed point problem inherently has $\varepsilon$ errors in every dimension, so we cannot ever expect to receive values *exactly* equal to 0 or 1. The two-player game gadgets in Chen, Deng and Teng (2009) and the market gadgets in Othman, Papadimitriou and Rubinstein (2016) also have this inherent limitation. Note that exact Boolean gates could be represented using a variant of Lemma 1 only once we have access to a comparison gate that produces an *exact* Boolean output value, and such a gate does not seem to be attainable for the previously-mentioned applications. Thus, using exact Boolean values would greatly diminish the applicability of the generalized circuits framework.

# 6 Conclusion

Generalized circuits are a vital tool for reasoning about the computational complexity of equilibrium approximation problems. In this paper, we have revealed a conceptual issue in the generalized circuits framework, namely that it lacks *monotonicity* of its approximate solution concept. We have shown that this creates subtle technical issues, including in prior work. To overcome these issues, we have shown that the Boolean gate types in these circuits are redundant features and that stronger Boolean gates can be defined based on the other (arithmetic and comparison) gates. We have shown that the resulting (equivalent) $\varepsilon$-GCIRCUIT$^{\text{SB}}$ problem satisfies monotonicity, serves as a drop-in replacement in prior work, and then eliminates the mentioned issues at a conceptual level. We have established monotonicity as a fundamental desideratum for any approximate solution concept.

Our results have implications for two potential future lines of research. First, future studies of generalized circuits (for example, hardness proofs for sub-classes of circuits) can consider either the $\varepsilon$-GCIRCUIT$^{\text{SB}}$ problem or the $\varepsilon$-GCIRCUIT$^{\text{NB}}$ problem, i.e., ignore Boolean gates altogether. Both of these variants satisfy monotonicity, which

---

graphical games that the Boolean gates are ultimately reduced to (via Lemma 1 and the game gadgets for the other gates) will be quite complicated. For those cases where a more direct representation is desired, we present a third PPAD-complete and monotonic variant of the $\varepsilon$-GCIRCUIT problem, called $\varepsilon$-GCIRCUIT$^{\beta}$, that allows for this. The parameter $\beta$ needs to be appropriately specified. See Appendix C. Note that our proof in Appendix C implies that the Boolean game gadgets in *do* actually satisfy the constraints for $\varepsilon$-GCIRCUIT, even though this is not shown in the paper. $\varepsilon$-GCIRCUIT is, however, not monotonic and the gadgets do not satisfy the stronger Boolean constraints in $\varepsilon$-GCIRCUIT$^{\text{SB}}$.

makes for a much more natural way of reasoning and avoids the kinds of technical pitfalls we have discussed. This may lead to new insights about computational complexity in generalized circuits. One such area of research are "support finding" problems, where we do not ask for numeric values, but only for a coarse discrete description of a solution. In our own recent work (Schuldenzucker, Seuken and Battiston, 2019, Section 5), we have studied one such PPAD-complete problem to prove hardness in the context of financial networks.

Another example where $\varepsilon$-GCircuit$^{SB}$ could be useful is a conjecture by Babichenko, Papadimitriou and Rubinstein (2016) that the following problem, termed $(\varepsilon, \delta)$-GCircuit, is already PPAD-complete for $\varepsilon, \delta \ll 1$: given a generalized circuit, find an assignment where the constraints for an $\varepsilon$-solution hold at least at a fraction of $1 - \delta$ of the gates. This would settle various open questions regarding the Nash equilibrium search problem. Given the benefits of monotonicity illustrated in this paper and towards a proof of the conjecture, it might be useful to instead consider the $(\varepsilon, \delta)$-GCircuit$^{SB}$ problem. Note that $(\varepsilon, \delta)$-GCircuit$^{SB}$ is monotonic in both parameters and $(\varepsilon, \delta)$-GCircuit$^{SB} \leq_P (\varepsilon/2, \Theta(\delta))$-GCircuit$^{NB}$ by Lemma 1.

The second strand of future research concerns reductions from generalized circuits to other problems to show PPAD-hardness of these problems. The redundancy of Boolean gates implies that no reduction needs to be provided for them, which will hopefully simplify these kinds of proofs in the future. Since the reduction now happens between two monotonic problems, their connection may further become more natural and allow for a more detailed study of common features.

# Appendix

## A  Examples for Issues #2–#4 in the Fan-Out 2 Reduction in Rubinstein (2018)

We present examples for issue #2–#4.

**Issue #2**  For issue #2, an example is given by our very first counterexample to monotonicity in Section 3.

**Issue #3**  For issue #3, assume that $v$ is the output of some arithmetic gate, let $g = C_\neg$ with input $v$, and assume that $v$ is input to at least two other gates so that its value needs to be distributed. Call this original circuit $C$ and let $C'$ be the circuit where a distribution subroutine is inserted after $v$. Let $v'$ be an output of the distribution subroutine and the new input to $g$ in $C'$. Let $w$ be the output of $g$ in $C$ and $C'$. Assume that there exists an $\varepsilon'$-solution $x'$ to $C'$ such that $x'[v] \geq 1 - \varepsilon'$, $x'[v'] \in (\varepsilon', 1 - \varepsilon')$, and $x'[w] = 0.5$. The distribution subroutine does not prevent this, no matter what $\varepsilon$ and $\varepsilon$', and it is easy to construct $C$ such that this actually happens. In $C$, we have for the input of $g$ that $x'[v] \geq 1 - \varepsilon' \geq 1 - \varepsilon$, but for the output $x[w] = 0.5$. So $x'$ is not an $\varepsilon$-solution for $C$. Note that this counterexample does not depend on the fact that $\varepsilon \neq \varepsilon'$.

**Issue #4**  For issue #4, consider a generalized circuit $C$ corresponding to the following definitions (where "=" assigns an output node to a gate):

$$z = 0.3$$
$$b = (a < z)$$
$$c = 1/2 \cdot b$$
$$d = 1/3 \cdot b$$
$$e = 1/4 \cdot b$$

Note that node $a$ is left unconstrained. We imagine that nodes $a$–$e$ are part of a larger circuit. Node $b$ has fan-out $3 > 2$, so Rubinstein's fan-out 2 reduction would attach a tree of double negation subroutines. The double negation subroutine is simply a chain of two negation gates connected by a new node. This turns an approximate TRUE into an approximate TRUE and an approximate FALSE into an approximate FALSE, but can return *any* value if its input is not approximately Boolean. The fan-out 2 reduction would now replace the definitions of nodes $c$–$e$ by the following to create a new reduced circuit $C'$:

$$b_1 = \neg\neg b$$
$$b_2 = \neg\neg b$$
$$b_{1,1} = \neg\neg b_1$$
$$b_{1,2} = \neg\neg b_1$$
$$b_{2,1} = \neg\neg b_2$$
$$c = 1/2 \cdot b_{1,1}$$
$$d = 1/3 \cdot b_{1,2}$$
$$e = 1/4 \cdot b_{2,2}$$

We now present a solution solution $x'$ to $C'$ that does not give rise to a solution to $C$. We will show that $c$ and $d$ can take on a combination of values in $C'$ that is not possible in $C$. Let $\varepsilon = 0.01$. Define $x'$ as follows:

$$x'[a] := x'[z] := 0.3$$
$$x'[b] := 0.5$$
$$x'[b_1] := 0.8$$
$$x'[b_2] := 0.2$$
$$x'[b_{1,1}] := x'[b_1] = 0.8$$
$$x'[b_{1,2}] := x'[b_1] = 0.8$$
$$x'[b_{2,1}] := x'[b_2] = 0.2$$
$$x'[c] := 1/2 \cdot x'[b_{1,1}] = 1/2 \cdot 0.8 = 0.4$$
$$x'[d] := 1/3 \cdot x'[b_{1,2}] = 1/3 \cdot 0.8 = 0.2\bar{6}$$
$$x'[e] := 1/4 \cdot x'[b_{2,1}] = 1/4 \cdot 0.2 = 0.05$$

For the interior nodes of the double negation subroutines, if the input node to the subroutine is $v$, set the interior node to value $1 - x'[v]$. This is always possible.

$x'$ is an $\varepsilon$-solution for $C'$. Note that, by choice of $x'[a]$, any value is allowed for $x'[b]$. We chose a value that is not approximately Boolean w.r.t. $\varepsilon$. That is why the following double negation subroutines can each output an arbitrary value at $x'[b_1]$ and $x'[b_2]$. The key to our counterexample is that these values need not be the same. The other gates then copy and transform the values normally.

$x'$ does not become an $\varepsilon$-solution for $C$ if we restrict it to nodes in $C$. That is because, if $x$ is any $\varepsilon$-solution to $C$, then $x[c] - x[e] = 1/2 \cdot x[b] \pm \varepsilon - 1/4 \cdot x[b] \pm \varepsilon = 1/4 \cdot x[b] \pm 2\varepsilon$. However, we have $x'[c] - x'[e] = 0.35 > 0.145 = 1/4 \cdot x'[b] + 2\varepsilon$.

Note further that i) the above value of $x'[c] - x'[e]$ would not be allowed in $C'$ for *any* value of $x'[b]$ and ii) we cannot guarantee the $\varepsilon$-solution property by increasing $\varepsilon$ by any constant factor.

# B  Minimal Modifications to Circumvent the Issues in Prior Work

We present a minimal set of modifications to Rubinstein (2018) and Chen, Deng and Teng (2009) that allow us to keep the current definition of the $\varepsilon$-GCircuit problem and that eliminate the problems discussed above. Our modifications are based on careful examination of the details of the involved proofs.

To show that issue 1–5 in Section 5 are not critical for the results of the respective papers, we exploit a common feature of the generalized circuit constructions from prior work, namely that Boolean gates do not occur at arbitrary positions. Their inputs always come from gates that are meant to yield approximately Boolean values, namely other Boolean gates and the comparison gate. Further, the output of each gate will be interpreted either as a Boolean value (by Boolean gates) or as a non-Boolean value (by other gates), but not both at the same time. Such circuits formally *still* do not satisfy monotonicity, but we can perform an additional normalization step after which they "essentially" do.

**Lemma 2** (Boolean-regular circuit)**.** *If $g$ and $g'$ are gates in a circuit such that the output of $g$ is an input to $g'$, then $g$ is called a* predecessor *of $g'$ and $g'$ is called a* successor *of $g$. We call a generalized circuit* Boolean-regular *if the following two conditions hold:*

1. *Any predecessor of any Boolean gate is either a Boolean gate itself or a comparison gate.*

2. *For any gate, if one of its successors is a Boolean gate, then all of its successors are Boolean gates.*

*If $C$ is Boolean-regular, then for any $\varepsilon$ and any $\varepsilon$-solution $x$ for $C$, we can compute in polynomial time an assignment $x'$ such that for any $\varepsilon' \geq \varepsilon$, $x'$ is an $\varepsilon'$-solution for $C$. We call an $x'$ resulting from this procedure* normalized*.*

*Proof.* Given $x$, define $x'$ as follows. If $v$ is not an input to any Boolean gate, then $x'[v] = x[v]$. If $v$ is an input to a Boolean gate, then

$$x'[v] = \begin{cases} 0 & \text{if } x[v] \leq \varepsilon \\ 1/2 & \text{if } x[v] \in (\varepsilon, 1-\varepsilon) \\ 1 & \text{if } x[v] \geq 1-\varepsilon. \end{cases}$$

Let now $\varepsilon' \geq \varepsilon$. We show that $x'$ is an $\varepsilon'$-solution. Let $g$ be any gate with inputs $a_1$ and $a_2$ (if any) and output $v$. We distinguish three cases.

- If $g$ is an arithmetic gate, then neither its output (by condition 1) nor any of its inputs (by condition 2) are input to any Boolean gate. Thus, $x' = x$ at these nodes. Since the constraints of arithmetic gates are monotonic in $\varepsilon$, the constraint at $g$ is still satisfied for $\varepsilon'$.

- If $g$ is a Boolean gate, then its constraints only distinguish the intervals $[0, \varepsilon']$, $(\varepsilon', 1-\varepsilon')$, and $[1-\varepsilon', 1]$. For each input $a_i$ of $g$, by definition of $x'$ it does not depend on $\varepsilon'$ to which of these three intervals $x'[a_i]$ belongs. Therefore, we require $x'[v] = 0 \pm \varepsilon'$ in an $\varepsilon'$-solution iff we require $x[v] = 0 \pm \varepsilon$ in an $\varepsilon$-solution. And the latter implies the former, both if $v$ is the input to another Boolean gate and if not. Likewise for $x'[v] = 1 \pm \varepsilon'$.

- If $g$ is a comparison gate, by condition 2 its inputs are not also input to any Boolean gate and thus $x'[a_i] = x[a_i]$ for $i = 1, 2$. Now we apply the same argument as for the outputs of Boolean gates to see that the constraint is still satisfied. □

Detailed examination of the proofs in the aforementioned two pieces of prior work shows that almost all construction steps lead to a Boolean-regular circuit. The only exception we are aware of is the EXTRACTBITS subroutine in Chen, Deng and Teng (2009), where the output of a $C_<$ gate is fed into both Boolean gates (which simulate a given Boolean circuit) and a $C_{\times \zeta}$ gate. Here, Boolean regularity can be easily restored by a minor modification to the construction.[15]

Towards issue #1 in Chen, Deng and Teng (2009), we can now WLOG consider the restriction of the $\varepsilon$-GCIRCUIT$_C$ problem where only Boolean-regular circuits are allowed as input and only normalized $\varepsilon$-solutions are allowed as output. Since this problem has monotonicity by definition of a normalized solution and the reduction in the proof of Theorem 5.7 in Chen, Deng and Teng (2009) preserves Boolean-regularity, issue #1 is eliminated.

Towards issue #2 and #3 in Rubinstein (2018), we notice that the fan-out 2 reduction preserves Boolean-regularity.[16] The restriction of $\varepsilon$-GCIRCUIT to Boolean-regular circuits and normalized solutions then resolves issues #2 (because it has monotonicity) and #3 (because the described situation does not occur by Boolean-regularity).

To see that issue #4 does not invalidate hardness of $\varepsilon$-GCIRCUIT restricted to fan-out 2, we again perform detailed examination of the arguments that are used in the main hardness proof. Issue #4 arises because the values at outputs of Boolean gates with non-Boolean input are allowed to be arbitrary and different in the reduced instance while they are arbitrary, but must be equal in the original instance (see our example in Appendix A). However, such a property is never exploited in the proof of hardness of the $\varepsilon$-GCIRCUIT problem. Instead, whenever the output of a Boolean gate can be arbitrary, it is accounted for as an independent $\pm 1$ error. Thus, if we apply the fan-out 2 reduction to the hard $\varepsilon$-GCIRCUIT instance, a solution to the reduced circuit is not necessarily a solution to the original circuit, but it *is* still a solution to the original hard BROUWER instance. And thus, the restriction to fan-out 2 is still hard.

Finally, to eliminate issue #5, one can study the Boolean game gadgets from Daskalakis, Goldberg and Papadimitriou (2009) to see that they in fact *do* satisfy the constraints for approximately Boolean values even though this is not stated explicitly in the paper. The proof is like in Proposition 6 in Appendix C, where we show it for $\varepsilon$-GCIRCUIT$^\beta$.

---

[15]One way to restore Boolean-regularity is to insert a double negation in front of the $C_{\times \zeta}$ gate. This will, of course, create additional degrees of freedom like in issue #4 (see Section 5). These are not a problem in this case for the same reason why issue #4, discussed below, is not critical.

[16]Here we assume WLOG that the trees of double negations are constructed in such a way that all outputs of the tree are all at the same level.

## C  A More Direct Representation of Boolean Gates

The way how we eliminated issue #5 may not feel very satisfying. When we perform reduction from $\varepsilon$-GCIRCUIT$^{\text{SB}}$ to other problems via Lemma 1, the representation of the Boolean gates will be rather indirect. Each Boolean gate is first represented by comparison gates, arithmetic gates, and using De Morgan's laws. Then *these* gates are represented as (say) game gadgets. In some situations, a more direct representation of Boolean gates may be desirable. This could be useful, for example, if one seeks to further modify the generalized circuit concept in a way incompatible with Lemma 1.

In this section, we present a way how such a direct representation of monotonic Boolean gates can be achieved. For our discussion, we focus on the reduction from generalized circuits to graphical games via the game gadgets in Daskalakis, Goldberg and Papadimitriou (2009). These are the same gadgets used in Rubinstein (2018). We will show that these game gadgets do not provide a reduction from $\varepsilon$-GCIRCUIT$^{\text{SB}}$. To overcome this, we will modify our solution concept again, which will lead to a new family of PPAD-complete search problems $\varepsilon$-GCIRCUIT$^\beta$, where $\beta \in (0, 1/2)$ is a parameter. We then show that the game gadgets provide a reduction from $\varepsilon$-GCIRCUIT$^\beta$ if $\beta$ is not too small. A drawback of this variant is that the $\beta$ parameter needs to be appropriately chosen for the individual application at hand.

Daskalakis, Goldberg and Papadimitriou (2009) and Rubinstein (2018) study binary graphical games in *$\varepsilon$-approximately well supported Nash equilibrium* (*$\varepsilon$-WSNE* for short). This means that players only have two actions, called 0 and 1, and if both strategies are played with positive probability in equilibrium, then the expected utilities from both pure actions must be $\varepsilon$-close to each other.[17] A mixed-strategy equilibrium of a binary game can be encoded by assigning to each player $i$ the probability $p[i]$ with which player $i$ plays action 1. *Game gadgets* are sub-games that in equilibrium enforce certain relationships, corresponding to the gates of a generalized circuit, on the $p[i]$ values of certain players.

The negation game gadget $\mathcal{G}_\neg$ (Daskalakis, Goldberg and Papadimitriou, 2009, Lemma 5.5) satisfies the constraints for a strong $\varepsilon$-solution, but the other two, $\mathcal{G}_\wedge$ and $\mathcal{G}_\vee$, do not. We consider $\mathcal{G}_\wedge$ in the following. The proof for $\mathcal{G}_\vee$ is analogous. Let $a$ and $b$ be two *input players* and let $v$ be an *output player*. The utility function of player $v$ in $\mathcal{G}_\wedge$ is defined as follows:

$$u_v = \begin{cases} 1/2 & \text{if } v \text{ plays } 0 \\ 1 & \text{if } v \text{ plays } 1 \wedge a \text{ plays } 1 \wedge b \text{ plays } 1 \\ 0 & \text{if } v \text{ plays } 1 \wedge (a \text{ plays } 0 \vee b \text{ plays } 0) \end{cases}$$

---

[17]Daskalakis, Goldberg and Papadimitriou (2009) prove that $\varepsilon$-WSNE and regular $\varepsilon$-approximate Nash equilibrium (where no deviation to any other mixed strategy can improve expected utility by more than $\varepsilon$) are equivalent if one scales $\varepsilon$ appropriately.

If player $v$ plays a pure strategy and the other players play mixed strategies according to $p$, the expected utility of $v$ is

$$\mathbb{E}\left[u_v\right] = \begin{cases} 1/2 & \text{if } v \text{ plays } 0 \\ p[a]p[b] & \text{if } v \text{ plays } 1. \end{cases}$$

This does not provide a reduction from $\varepsilon$-GCircuit$^{\text{SB}}$, no matter how much we reduce $\varepsilon$ in the transition from generalized circuits to games:

**Proposition 5.** *There exists an $\varepsilon > 0$ such that there is no $\varepsilon' > 0$ such that, whenever $\mathcal{G}_\wedge$ occurs as part of a larger game and $p$ is an $\varepsilon'$-WSNE, $x := p$ satisfies the constraints for $C_\wedge$ for a strong $\varepsilon$-solution.*

*Proof.* Consider an $\varepsilon$ for which such an $\varepsilon'$ *does* exist. Let $p[a] = p[b] = 1/2 + 2\varepsilon$. Then Table 2 prescribes that $p[v] \geq 1 - \varepsilon$. To guarantee *any* statement of form "$p[v] \geq ...$" in an $\varepsilon'$-WSNE, we require

$$\left(\frac{1}{2} + 2\varepsilon\right)^2 = p[a]p[b] = \mathbb{E}\left[u_v\right](1, p_{-v}) > \mathbb{E}\left[u_v\right](0, p_{-v}) = 1/2 + \varepsilon'.$$

By simple algebra, this implies that

$$\varepsilon > \frac{1}{24} + \frac{1}{6}\varepsilon' > \frac{1}{24}.$$

Therefore, for $\varepsilon \leq \frac{1}{24}$, we can always choose $p[v] = 0$ even though the constraints for a strong $\varepsilon$-solution prescribe $p[v] \geq 1 - \varepsilon$. Thus, $x := p$ is not a strong $\varepsilon$-solution. $\square$

The previous proposition shows that the game gadgets in Daskalakis, Goldberg and Papadimitriou (2009) do not imply sufficiently strong constraints to imply a direct representation of the Boolean gates in $\varepsilon$-GCircuit$^{\text{SB}}$. However, we can make the solution concept itself slightly weaker to accommodate these gadgets while preserving monotonicity and hardness.

To do this, let $\beta < 1/2$ and $\varepsilon < \beta, 1/2 - \beta$. Given a generalized circuit, we call an assignment $x$ an $\varepsilon^\beta$-*solution*[18] if it satisfies the constraints in Table 2 where we replace $\varepsilon$ by $\beta$ in the preconditions of all Boolean gates. That is, for the Boolean gates we have the constraints in Table 3. We call the corresponding search problem $\varepsilon$-GCircuit$^\beta$.[19]

---

[18]We chose this notation to avoid confusion with the (unrelated) concept of an $(\varepsilon, \delta)$-solution in Babichenko, Papadimitriou and Rubinstein (2016), where a $1 - \delta$ fraction of constraints needs to be satisfied up to precision $\varepsilon$.

[19]Our definition of $\varepsilon$-GCircuit$^\beta$ is inspired by Othman, Papadimitriou and Rubinstein (2016), where we however do *not* consider $\beta = \Theta(\varepsilon)$, but $\varepsilon \ll \beta$. Note further that we do not use $\beta$ in the precondition of the comparison gate. This would make for an even weaker problem, but a *too* weak one: Rubinstein's (2018) hardness proof performs comparison with multiples of $\sqrt{\varepsilon}$ and the "brittleness" of the comparison gate needs to be significantly smaller than that.

**Table 3** Constraints for the Boolean gates in an $\varepsilon^\beta$-solution. All other constraints are the same as in a (strong) $\varepsilon$-solution.

| Gate Type | Short | Constraint |
|---|---|---|
| OR | $C_\vee$ | $x[a_1] < 1/2 - \beta$ and $x[a_2] < 1/2 - \beta \Rightarrow x[v] \leq \varepsilon$ |
|  |  | $x[a_1] > 1/2 + \beta$ or $\ \ x[a_2] > 1/2 + \beta \Rightarrow x[v] \geq 1 - \varepsilon$ |
| AND | $C_\wedge$ | $x[a_1] < 1/2 - \beta$ or $\ \ x[a_2] < 1/2 - \beta \Rightarrow x[v] \leq \varepsilon$ |
|  |  | $x[a_1] > 1/2 + \beta$ and $x[a_2] > 1/2 + \beta \Rightarrow x[v] \geq 1 - \varepsilon$ |
| NOT | $C_\neg$ | $x[a_1] < 1/2 - \beta \Rightarrow x[v] \geq 1 - \varepsilon$ |
|  |  | $x[a_1] > 1/2 + \beta \Rightarrow x[v] \leq \varepsilon$ |

The two parameters $\varepsilon$ and $\beta$ take on different roles. Typically, $\beta$ will be fixed to an arbitrary, not necessarily small, constant, like $1/4$. Then $\varepsilon$ is chosen arbitrarily small. It is easy to see that the solution concept is monotonic in both parameters and that, for any fixed $\beta$ and sufficiently small $\varepsilon$ depending on $\beta$, any strong $\varepsilon$-solution is an $\varepsilon^\beta$-solution and any $\varepsilon^\beta$-solution is an $\varepsilon$-solution. This immediately implies that $\varepsilon$-GCIRCUIT$^\beta$ is PPAD-complete for any $\beta < 1/2$ and $\varepsilon \ll 1$ depending on $\beta$.

The Boolean game gadgets satisfy the constraints for an $\varepsilon^\beta$-solution, and thus define a reduction from $\varepsilon$-GCIRCUIT$^\beta$, as long as $\beta$ is not too small.

**Proposition 6.** *Let $1/4 < \beta < 1/2$ and let $\varepsilon \leq \beta - 1/4$. Let $o \in \{\vee, \wedge, \neg\}$ and consider the binary graphical game $\mathcal{G}_o$ from Daskalakis, Goldberg and Papadimitriou (2009, Lemma 5.5) with input players $a$ and $b$ (if any) and output player $v$. Then any $\varepsilon$-WSNE $p$ satisfies the constraint corresponding to the gate $C_o$ and $x := p$ for an $\varepsilon^\beta$-solution.*

*Proof.* We show the statement for $o = \wedge$. The other operations are similar. Assume that $p[a] > 1/2 + \beta$ and $p[b] > 1/2 + \beta$. Then $\mathbb{E}[u_v](1, x_{-v}) = p[a]p[b] > (1/2 + \beta)^2 > 1/2 + \varepsilon = \mathbb{E}[u_v](0, x_{-v}) + \varepsilon$, where the middle inequality is by choice of $\beta$ and $\varepsilon$. Since we are in an $\varepsilon$-WSNE, this implies $p[v] = 1$ and in particular $p[v] \geq 1 - \varepsilon$.

Vice versa, assume that $p[a] < 1/2 - \beta$ or $p[b] < 1/2 - \beta$. Then $\mathbb{E}[u_v](1, x_{-v}) = p[a]p[b] < 1/2 - \beta < 1/2 - \varepsilon = \mathbb{E}[u_v](0, x_{-v}) - \varepsilon$ and thus $x[v] = 0 \leq \varepsilon$ by the $\varepsilon$-WSNE. □

By more careful analysis of the error terms in the previous proof, one can show that the gadget still works for all $\beta > (\sqrt{2} - 1)/4 \approx 0.10$ and $\varepsilon \ll 1$ (the threshold for $\varepsilon$ depending on $\beta$), but not for smaller $\beta$.